\documentstyle[NATO]{crckapb}
\input epsf.sty
\begin{opening}
\title{Theories of Hydrophobic Effects and The Description of Free 
Volume in Complex Liquids}

\author{LAWRENCE R. PRATT, SHEKHAR GARDE, AND GERHARD HUMMER}
\institute{Theoretical Division, Los Alamos National Laboratory 
\\Los Alamos, New Mexico 87545 USA }
\end{opening}

\runningtitle{Free Volume and Hydrophobic Effects}

\begin{document}

\begin{abstract}
Recent progress on molecular theories of hydration of nonpolar solutes
in aqueous solution has led to new ways of thinking about the old
issue of free, or ``available'', volume in liquids.  This article
surveys the principal new results with particular attention to general
issues of packing in liquids.

\end{abstract}

\section{Introduction}
Aqueous solutions of colloidal solutes are preeminent examples of
complex liquids.  In such settings, attention is often directed
towards the issues of macromolecular structure, aggregation, and
dynamics.  However, one aspect of these problems is stubbornly
associated with the small size scale of molecules, in particular the
small size of a water molecule.  That problem is the molecular
understanding and description of hydration effects -- hydrophobic
effects -- that stabilize membranes, micelles, folded proteins, and
aggregates of such structures.  This presentation will focus on that
basic, molecular scale issue underlying aqueous solutions of interest
both to biophysics and material science of colloids.

Because hydrophobic effects are so broadly discussed, it might be
surprising to notice that workers on molecular theories of hydrophobic
effects have not achieved good agreement on that molecular theory, at
least when we proceed beyond the primitive stage of the principles
of statistical mechanics and specification of the intermolecular
interactions involved.  The empirical fact is that different
researchers hold different opinions on the correctness of several
available theories, each of which agrees with a set of experimental
data at least roughly.  A sampling of different perspectives is
available from the references [1-33];
%\cite{Ben-Naim:67,Stillinger:73,Lee:85,Bennaim,Privalov:89,%
%Pohorille:90,Dill:90,Privalov:90,Murphy:90,Muller,Pratt:91,Sharp:91a,%
%Sharp:91b,Pratt:92,Lazaridis:92,Pratt:93,Lee:93,Belle:93,Ben-Naim:93,%
%Holtzer:94,Dang:94,Belle:94,Madan,Sitkoff,Holtzer:95,Kumar:95,%
%Horvath,Headgordon,Haymet,Garde:96a,Sharp:96,Prevost:96,Wiggins}; 
this collection is not intended to be complete, however.

Furthermore, the simplest extensions of our experimental information
can spark new debates about our understanding of hydrophobic effects.
A current example, the effect of pressure on hydrophobic stabilization
of folded proteins, is discussed below.

In consequence of such observations, the theory surveyed here was
designed for maximal simplicity on the fundamentals of statistical
mechanics and on the physical assumptions applied to the particular
problems.  The theory that was developed is relevant to the
traditional packing problems of theories of liquids and we emphasize
that connection in the discussion here.

A central problem for the theory of hydrophobic effects is the old,
but imperfectly solved, theoretical problem of finding space for a
solute in liquid solvents.  As is well recognized, liquids are dense,
disordered materials. It is this combination of attributes that makes
these problems difficult.  The description of available volume in hard
core model liquids is central to the modern understanding of the van
der Waals equation of state and is basic to such ordering phase
transitions as the hard sphere freezing and the liquid crystal phase
transitions \cite{WCA:83}. In the specific motivating case considered
here, the particular molecular structuring characteristic of liquid
water is expected to be important.  So we must preserve the fidelity
of the description of the structure of liquid water, in addition to
tackling the problem of packing in dense, disordered materials.

This lecture develops a new conceptualization of this old problem and
new techniques for predicting the fractional free volume accessible to
hard core molecules in condensed phases.  This new approach is based
upon an information theory perspective that has general applicability
and was initially explicitly heuristic.  Information was sought on a
condensed medium of interest and on that basis a prediction of the
fractional free volume accessible to a hard model solute was made. For
some important cases considered so far, the detail of information
required for accurate, interesting predictions has been surprisingly
modest.

\section{Available Volume Statistics}
Consider the solubility of inert gases in aqueous solutions.  The
medium is liquid water and the solute is idealized as a hard object,
perfectly repelling the center (oxygen atom) of each water molecule.
For such models the interaction part of the chemical potential of the
solute is obtained as
\begin{equation} \beta\Delta\mu=-\ln p_0~, \label{eqn:ktlnp0}
\end{equation} with p$_0$  the probability that the hard solute
could be inserted into the system without overlap of van der Waals
volume of the solvent; 1/$\beta$=k$_B$T.  This is a specialization of
Widom's formula \cite{Widom:63,Widom:82}
\begin{equation}
\exp\{-\beta\Delta\mu \} = \langle \exp\{-\beta\Delta U\}
\rangle_0~.
\label{eqn:widom} 
\end{equation}
$\Delta U$ is the change in the solute-solvent interaction potential
energy upon placement of the solute in an arbitrary position in the
solvent and the average indicated by $\langle \ldots \rangle_0$ is
over the thermal motion of the solvent unaffected by the solute.  The
solute is a test particle for this calculation.  For the hard core
model being considered, $\Delta U$ is either {\em zero} or {\em
infinity}, so the average sought involves a random variable with value
either {\em one} or {\em zero}; the averaging collects the fraction of
solute placements that would be allowed.  If presented with a thermal
configuration of a large volume of solvent, we might estimate these
quantities by performing many trial placements of the solute
throughout the solvent and determining the fraction of those trial
placements that would be allowed.  This estimates $V_{free}/V$, the
fractional free volume accessible to the solute.  Thus,
Eq.~(\ref{eqn:ktlnp0}) is a free volume formula \cite{Reiss:92}, exact
for the model being considered.

The operation of these formulae can be viewed alternatively: Imagine
identifying a molecular scale volume at an arbitrary position in the
liquid system by (1) hypothetical placement of the solute and (2)
determination of those positions of water oxygen atoms that would be
excluded due to solute-solvent interactions.  We will call this volume
the observation volume.  With such a molecular scale volume defined we
could keep track, say during a simulation calculation, of the
probabilities p$_n$ that $n=0, 1,
\ldots$ oxygen atom occupants are observed.  As the notation suggests,
p$_0$ is the probability that no occupants are observed in the
molecular volume.  

Our strategy for predicting of p$_0$ will be to model the distribution
p$_n$ and to extract the extreme value p$_0$.  This is a primitive
approach to theories of $\beta\Delta\mu$ and solubilities of inert
gases water.  Both more and less subtle theoretical works on these
topics have been long available.  ``Less subtle'' here means
simulation calculations, techniques more straightforwardly useful than
many ``more subtle'' approaches.  The ``more subtle'' means here that
further statistical quantities have been introduced for spherical
solutes with the intention that they might facilitate more expansive
approximate theories.  These include
\begin{eqnarray}
d_1(\lambda)= -{d p_0(\lambda)\over d \lambda}~,
\label{eqn:dfn.a}
\end{eqnarray}
\begin{eqnarray}
G(\lambda) = ({-1\over 4 \pi
\lambda^2 \rho}){d \ln p_0(\lambda)\over d \lambda}~.
\label{eqn:dfn.b}
\end{eqnarray}
$\lambda$ is the radius of a center-to-center exclusion sphere.  These
quantities have been useful in suggesting physical theories because
they have interpretations that are appreciated physically.
$d_1(\lambda)$, is the distribution function of distances $\lambda$
from an arbitrary point in the liquid to the {\em nearest} solvent
center. $4\pi\lambda^2\rho G(\lambda)$ [Eq.~(\ref{eqn:dfn.b})] is, in
view of Eq.~(\ref{eqn:ktlnp0}), the derivative with respect to
exclusion radius of the hydration free energy due to intermolecular
interactions, in thermal energy units k$_B$T.  It gives the
compressive force exerted by the solvent on the hard spherical solute.
In addition, $4 \pi\lambda^2 \rho G(\lambda)d\lambda$ is the expected
number of solvent centers in a shell of radius $\lambda$ and width
$d\lambda$ outside a hard sphere that excludes solvent centers from a
ball of radius $\lambda$.

\section{Simulation Results for Liquid Water}
Some of the simulation work has determined the quantities of
Eqs.~(\ref{eqn:dfn.a}) and (\ref{eqn:dfn.b}) for molecular liquids
represented realistically at the current state-of-the-art
\cite{Pratt:92,Pratt:93}. Thus we know that $d_1(\lambda)$ for liquid
water and for liquid n-hexane are both unimodal with maximum displaced
by a distance less than 0.1\AA; the maximum occurs at slightly smaller
distances for liquid n-hexane than for liquid water
\cite{Pratt:92,Pratt:93}. This difference in the most probable cavity
size between liquid n-hexane and liquid water is not large.  The
difference in the most probable cavity size between liquid water and a
reference random medium, with sites of the same radius and distributed
randomly at the water density, is greater than the difference between
n-hexane and water.  So the differences observed between the two
molecular liquids considered are not {\em purely} reflections of
molecular size and density.  This comparison addresses the idea that
the low solubility of inert gases in liquid water might be due to the
small size of the water molecule and the possibility that
``interstitial'' cavities would, on this basis, be smaller in water
than in the organic liquid \cite{Lee:85,Lee:91}. The fact that these
differences between water and n-hexane are slight is associated with
the fact that the basic units considered in n-hexane are the methyl
and methylene groups.  These are not so different in size from a water
molecule. It should be noted also that on a packing fraction basis,
typical organic liquids are denser than liquid water
\cite{Pohorille:90}.

The notable distinction between the results for liquid water and
liquid n-hexane is that the distribution $d_1(\lambda)$ is narrower
for liquid water.  This suggests that the liquid water phase is less
flexible than the liquid n-hexane phase in opening cavities of
substantial size.

Simulation calculations have also produced $G(\lambda)$ for
0$<\lambda<$3.0\AA, approximately.  This size range covers
the simplest atomic solutes He and Ne but not much more.  This does,
however, permit comparison between water and organic solvents, and it
permits comparison of available theories with the simulation data.
Thus for the range 2.0\AA$<\lambda<$3.0\AA, $G(\lambda)$
for liquid water is approximately two-times larger than for n-hexane.
Water exerts a higher compressive force on the surface of an inert
solute than do typical organic liquids; water squeezes-out hydrophobic
solutes \cite{Richards:SA:91}.

The checking of theories against the available simulation data
$G(\lambda)$ for water has also been revealing
\cite{Pratt:92,Pratt:93}. We now know that the predictions of the scaled 
particle model \cite{Pierotti:63,Pierotti:76} are significantly below
the numerically exact results for $G(\lambda)$. The Pratt-Chandler
(PC) integral equation theory [42-44] predicts results for
$G(\lambda)$ that are significantly too large.  The more pragmatic
revised scaled particle model due to Stillinger
\cite{Stillinger:73} typically predicts $G(\lambda)$ between those two
theories and with some empiricism about interpolation junctions can
describe the available simulation data satisfactorily
\cite{Pratt:92,Pratt:93}. For sizes $\lambda\gg$3.0\AA~the available
computer simulation data are less extensive, the theories less
convincing, and the checking has been pursued less vigorously.  See,
however, the recent results of Reference~\cite{Hummer:PRL:98}.
\begin{figure} [hbp]
\begin{center}
\leavevmode
\epsfbox{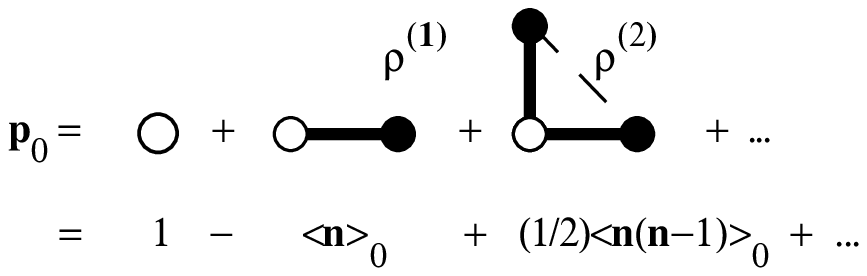}
\end{center}
\caption{Mayer-Montroll expansion for the insertion probability p$_0$.
The standard notation follows \protect{\cite{Andersen:77,Hansen}},
{\em e.g.}, the solid bonds indicate Mayer f-functions.  The second
line gives the evaluations for the diagrams shown in the case of a
hard core solute.  $n$ is the number of solvent centers in the
observation volume.%  See Eq.~(\protect{\ref{eq:facmom}}).
}
\label{fig:MM} \end{figure}

\section{Information Model}
We now return to the theoretical program of predicting p$_0$. What are
the standard theoretical tools for this?  The most immediate guiding
theory is the `inclusion-exclusion' development \cite{Riordan:78} of
Eq.~(\ref{eqn:widom}) \cite{Mayer:41,Reiss:59,vanKampen}:
\begin{eqnarray}
  \label{eq:pzero_corr} p_0 & = & 1 + \sum_{m=1}^\infty \frac{(-1)^m
}{m!} \int_v d{\bf r}_1 \int_v d{\bf r}_2 \cdots \int_v d{\bf r}_m
\rho^{(m)}({\bf r}_1, {\bf r}_2, \ldots, {\bf r}_m)~,
\end{eqnarray}
where $\rho^{(m)}$ is the $m$-body joint density for solvent centers.
This is depicted in a standard way in Figure~\ref{fig:MM}.  These are
standard combinatorial results, frequently seen in forms such as
\cite{Riordan:78}
\begin{eqnarray}
  \label{eq:facmom} p_0 & = & 1 -\langle n\rangle_0 + \sum_{m=2}^\infty
\frac{(-1)^m }{m!} \langle{n(n-1)\cdots(n-m+1)}\rangle_0~.
\end{eqnarray}
Here the random variable $n$ is the number of solvent centers within
the observation volume and, {\em e.g.}, $\langle n \rangle _0$ is the
expected number of centers within the observation volume.  Several
important points can be made from Figure~\ref{fig:MM}.  The first is
that the  primordial available volume
model is obtained from the first two terms shown
$\beta\Delta\mu\approx - \ln [ 1-\langle n \rangle _0 ]$.  The second
point is direct and basic: p$_0$ is naturally expressed in terms of
occupancy moments, indeed binomial moments here.  The sum truncates
sharply for cases where a finite maximum number of particles can be
present in the observation volume.  The sum can be of practical value
in the case where only the first nontrivial term is
retained.  For large solute volumes or solvent densities, this sum is
not directly useful; we seek a way to exploit the same information but
in a more broadly useful form.

\begin{figure}
\begin{center}
\leavevmode
\epsfbox{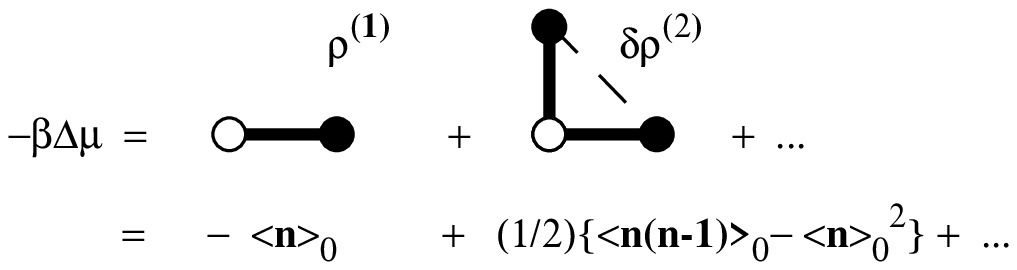}
\end{center}
\caption{`Virial' expansion, notation as in 
Figure~\protect{\ref{fig:MM}}.  Virial typically connotes a low
density expansion but here we are assuming full knowledge of the
medium correlation functions in the absence of the solute;
$\delta\rho^{(2)}(1,2)=\rho^{(2)}(1,2)-\rho^{(2)}(1)\rho^{(1)}(2)$.
Thus from the perspective of a density expansion this organization
shuffles the contributions to the virial coefficients.  In fact, here
contributions are ordered according to the number of bonds attached to
the root point.  Successive contributions have a structure that may be
derived from that of familiar cumulants with the formal replacement
$\langle n^k \rangle_0 \rightarrow \langle n(n-1)\cdots(n-k+1)
\rangle_0$; then p$_0$ can be formally expressed as 
$\langle e^{-n} \rangle_0$.
See~\protect{\cite{Kubo,Carruthers:91,Knuth}}.  Table~1 gives formulae
for contributions through 5th order.}
\label{fig:virial} \end{figure}

The next most immediate theoretical guidance comes from the `virial'
expansion depicted in Figure~\ref{fig:virial}.  This can be considered
a resummation of the series Figure~\ref{fig:MM} and is better in the
sense that such a truncation cannot produce a negative probability as
truncation of Figure~\ref{fig:MM} can do.  However, when that trouble
is avoided Figure~\ref{fig:virial} can still be less compact and
truncations can be less accurate.  For example, 
single term approximation to Figure~\ref{fig:MM} can be satisfactory
and then Figure~\ref{fig:virial} with a single term is likely to be
less so.

\begin{table}[htb]
\begin{center}
\caption{Successive contributions to the series 
$\beta \Delta \mu =\sum_1^\infty a_j$ of
Figure~\protect{\ref{fig:virial}}. \hfill ${n\choose
j}\equiv n(n-1)\cdots(n-j+1)/j!$ is the binomial coefficient.}
\begin{tabular}{cc}
\hline
j  &  a$_j$  \\ 
1 & $+\langle {n \choose 1}\rangle_0{}$ \\ 
2 & $
-\langle {n \choose 2} \rangle_0
+ \langle {n \choose 1} \rangle_0{}^2 /2$ \\

3 & $
+\langle {n \choose 3} \rangle_0 
- \langle {n \choose 1}\rangle_0\langle {n \choose 2} \rangle_0
+ \langle {n \choose 1} \rangle_0{}^3 /3$ \\

4 & $
-\langle {n \choose 4} \rangle_0 
+\langle {n \choose 1} \rangle_0\langle {n \choose 3} \rangle_0 
+ \langle {n \choose 2} \rangle_0{}^2 /2
- \langle {n \choose 1} \rangle_0{}^2\langle {n \choose 2} \rangle_0 
+ \langle {n \choose 1} \rangle_0{}^4 /4$ \\

5  & $
+\langle {n \choose 5} \rangle_0 
- \langle {n \choose 1} \rangle_0\langle {n \choose 4} \rangle_0 
+\langle {n \choose 1} \rangle_0{}^2\langle {n \choose 3} \rangle_0$
\protect{\hfill} \\
 & \protect{\hfill}$
- \langle {n \choose 2} \rangle_0\langle {n \choose 3} \rangle_0  
+\langle {n \choose 1}\rangle_0\langle {n \choose 2}\rangle_0{}^2
-\langle {n \choose 1}\rangle_0{}^3\langle {n \choose 2} \rangle_0
+\langle {n \choose 1} \rangle_0{}^5/5$  \\ 
\hline
\end{tabular}
\end{center}
\end{table}
 
Ultimately these considerations avoid the issue that we have only
limited information and we want to make the best prediction of p$_0$
that we can.  When the problem is stated this way what to do next is
clear: we model the probabilities $p_n$ on an information theory
basis.  We consider a relative or cross information entropy
\cite{Shore:80},
\begin{eqnarray}
  \eta(\{p_n\}) & = & - \sum_{n=0}^{\infty} p_n \ln
  \left(\frac{p_n}{\hat{p}_n}\right)~,
  \label{eq:entropy}
\end{eqnarray}
where $\hat{p}_n$ represents a ``default model'' chosen heuristically.
The moments that enter into the series Figure~\ref{fig:MM},
obtained from simulations if necessary, are the information typically
used.  We then maximize this information entropy subject to the
constraints that the probabilities reproduce the available
information.  The formal maximization of this entropy gives
probabilities
\begin{equation} p_j \propto {\hat p}_j  \exp
\left(-\sum_{k=1}^{k_{max}} \zeta_k {j \choose k} \right)  \label{maxpj}
\end{equation} 
where the $\zeta_k$ are Lagrange multipliers to be adjusted so that
the probabilities finally reproduce the information given initially.
The machinery for doing this can be developed straightforwardly.  For
example, the normalization of the probabilities can be deferred at
intermediate stages of the calculation.  Then the final thermodynamic
result can be given in terms of the required normalization factor
\begin{eqnarray}
 \beta\Delta\mu = \ln \sum_{n=0}{\hat{p}_n \over \hat{p}_0} \exp
\left(-\sum_{k=1}^{k_{max}} \zeta_k {n \choose k} \right)~.
\label{eq:pf}
\end{eqnarray} 
This is suggestive of the calculation of a partition
function for a modest-sized set of states with effective interactions.
The interesting questions then involve the predictions extracted from
the p$_j$ for properties other than the given information.  In our
case, the property of first interest is $\beta\Delta\mu$.

We can make the comforting observation that use of only
$\langle n\rangle_0$ and the natural default model ${\hat p}_j
\propto$ 1/j! produces the Poisson distribution as expected; $\zeta_1
= -\ln\langle n\rangle_0$ and $\beta\Delta\mu = \langle n\rangle_0$ as
would be found by retaining only the first term shown in
Figure~\ref{fig:virial}.  Note that evaluation of the second term
there with the Poisson distribution gives zero as it should
\cite{Carruthers:91}; $\langle\delta n^2\rangle_0=\langle n\rangle_0$
for the  Poisson distribution.
\begin{figure} [hbp]
\begin{center}
\leavevmode
\epsfxsize = 3.6in
\epsfysize = 2.6in
\epsfbox{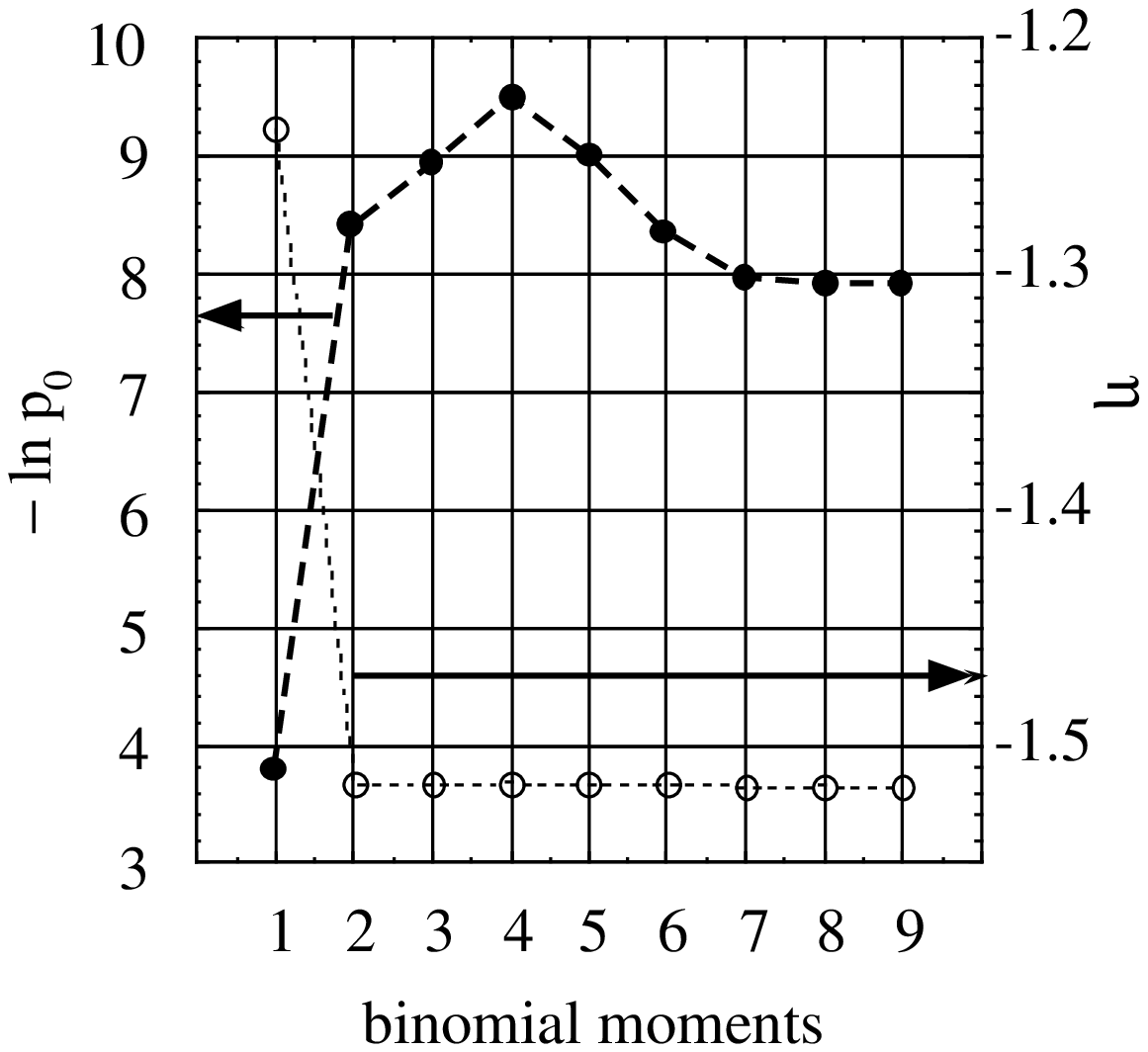}
\end{center}
%\hspace{0.5cm}%
%\psfig{figure=convergence.eps,width=4.5in,height=3.2in,angle=0}
\caption{Convergence of predictions for hydration free energy and 
information entropy with numbers of binomial moments employed; ${\hat p}_j
\propto$ 1/j!.  The
moment information was obtained from computer simulation of liquid
water \protect{\cite{Hummer:96}}. $\lambda$=3.0\AA.  The filled (open)
circles use the left (right) scale.}
\label{fig:convergence} \end{figure}

Figure~\ref{fig:convergence} shows how this prediction scheme
works-out when the solvent is computer simulated liquid water and the
solute is a hard sphere of a size appropriate for comparison with a Ne
atom.  The immediate point is that a model based upon the two moments
that might be obtained from experiment, $\langle n\rangle_0$ and
$\langle n(n-1)\rangle_0$, fortuitously provides the most satisfactory
simple prediction of p$_0$.  The Poisson (one moment) model is not
satisfactory.  Inclusion of moments higher than the second is not
advantageous unless several higher moments are available.  The
behavior of the information entropy $\eta$ suggests that the initial
two moments do a good job of describing the distribution and the
subsequent higher moments are `uninformative.'

These points are further remarkable because recent analyses
\cite{Chandler:93,Percus:93} have underscored the fact that
Percus-Yevick approximate integral equations can be derived on the
assumption that solvent density fluctuations are distributed according
to a Gaussian probability functional.  The results of the present
investigations, including particularly [56-59], give further insight
and support to those ideas.  The PC approximate integral equation
theory of hydrophobic effects \cite{Pratt:77}, at its inception a
Percus-Yevick analogue, can be given a similar basis
\cite{Chandler:93}.  The PC theory is thereby given a better
foundation than was available at its genesis.

Note that present two-moment model is not precisely an assumption of a
Gaussian probability functional for a density field.  The occupancies
here are required to be nonnegative integers.  That this be true of {\em
all} subvolumes of the observation volume is an important restriction.

\begin{table}[htb]
\begin{center}
\caption{Approximate evaluation of  contributions to the series 
Figure~\protect{\ref{fig:virial}} for the circumstances of
Figure~\protect{\ref{fig:convergence}}, $\lambda$=3.0\AA.  The final
row gives the result obtained directly from simulation
\protect{\cite{Hummer:96}}.  The two right-most columns 
elaborate a direct idea for reorganization of this series, motivated
by the form of the single-term result $\beta\Delta\mu\approx - \ln [
1-\langle n \rangle _0 ]$. Though this resummation improves the
prediction after each term, after five terms the result is still not
as successful as the two-moment information theory shown in
Figure~\protect{\ref{fig:convergence}}.  }
\begin{tabular}{cccccc}
\hline
j & $\langle {n \choose j} \rangle_0$ & a$_j$ &
$\sum_1^j a_{k}$ &  $ r_j\equiv{a_j/a_{j-1}\over(1-1/j)}$
& $-\ln[1-r_j]+\sum_1^j [a_k-r_j{}^k/k]$ \\ 
1 & 3.77 & 3.77 & 3.77 & - & - \\
2 & 5.75 & 1.36 & 5.13 & 0.72 & 5.42 \\ 
3 & 4.57 & 0.75 & 5.88 & 0.83 & 6.29 \\ 
4 & 2.04 & 0.50 & 6.38 & 0.89 & 6.91 \\
5 & 0.51 & 0.36 & 6.74 & 0.90 & 7.21 \\
$\infty$ & - & - & 7.93& -  & - \\
\hline
\end{tabular}
\end{center}
\end{table}

That specifically binomial moments are involved in the series above
emphasizes the point that the occupancies must be nonnegative
integers.  Further perspective on the convergence issue is obtained by
examination of successive contributions to the series
Figure~\ref{fig:virial}.  Numerical results are shown in Table~2.  The
first and second order terms make a significant contribution but by
themselves are not close to the full answer.  The values of the
additional terms do not establish a rapid convergence to the known
full answer.  However, we can view the information theory model as a
technique for reorganization of the series.

\subsection{Entropy Convergence}
When the typical occupation numbers $n$ are large the granularity of
the distribution p$_n$ is expected to be less significant, at least
near the center of the distribution and when viewed on a coarse enough
scale.  In such circumstances the predictions of two-moment
information theory models are not significantly different than those
of the PC theory.  It is remarkable that a simple calculation along
these lines gives a convincing explanation of the puzzling and
contentious issue of ``entropy convergence'' in hydrophobic
hydration \cite{Garde:96}.

The phenomenon to be explained is the following: entropies of transfer
of non-polar molecules from gas phase or a non-polar solvent into
water converge at a temperature of about 400~K to approximately zero
entropy change.  Similar behavior was also seen in the
microcalorimetry experiments on unfolding of several globular
proteins.  This behavior is insensitive to the particular hydrophobic
solute molecule.  Since the entropy is a temperature derivative of a
hydration free energy, the convergence temperature identifies a region
where graphs of hydration free energy versus temperature are extremal,
in fact, maximal.  Below that region the hydrophobic hydration free
energy increases with temperature but above that region the
hydrophobic hydration free energy decreases as the temperature is
raised.

The two-moment information theory model above was applied to this
problem for hard sphere solutes in water with the heuristic
modification that a flat default model was used; ${\hat p}_j
\propto$ constant for j$\le$j$_{max}$ and zero otherwise.  This latter
adjustment was found empirically to give slightly better hydration
free energies.  The results of the model and simulation calculations
accurately agreed on the temperature dependence of the hydration free
energies.  To analyze this agreement the information theory model was
simplified to a continuous Gaussian distribution that then gives
\begin{eqnarray}
\Delta\mu &\approx& 
{k_{\rm B}T\over 2}\left\{ {\langle n \rangle_0{}^2 \over 
\langle \delta n^2 \rangle_0} + 
\ln[2\pi \langle \delta n^2 \rangle_0] \right\} 
\label{eq:muex.a} \\ 
	& = & T\rho_{sat}(T){}^2\{k_{\rm B} v^2/2\langle \delta n^2
\rangle_0\} + T \{k_{\rm B}~\ln(2\pi \langle \delta n^2 \rangle_0]/2
\}~.
\label{eq:muex.b} \end{eqnarray}
$v$ is the observation volume and $\rho_{sat}(T)$ is the liquid
density along the vapor saturation curve so that $\langle n
\rangle_0 = \rho_{sat}(T)v$. 
$\langle \delta n^2 \rangle_0$ was found to be insensitive to
temperature for the relevant conditions.  Further, the first term of
Eq.~(\ref{eq:muex.b}) is larger than the second.  Thus the
non-monotonic behavior of the free energy with temperature and the
entropy convergence is a consequence of the non-monotonic variation of
$T\rho_{sat}(T){}^2$ with temperature.  The only molecular parameter
to complicate matters is the volume $v$ and with this formula $v$ does
not affect the entropy convergence temperature
\cite{Garde:96}. Thus the temperature of entropy convergence is about
the same for a wide family of solutes.

The physical point is: the entropy convergence phenomenon occurs
for water because of the low and temperature insensitive values of
$\langle \delta n^2 \rangle_0$.  In fact, the isothermal
compressibility of water at low pressure has a minimum value at T=319
K.  That temperature differs substantially from the observed entropy
convergence temperatures but it is not necessary that these
temperatures be approximately equal, just that $\langle \delta n^2
\rangle_0$ be insensitive to temperature in the region $d
[T\rho_{sat}(T){}^2]/dT\approx$0.

The technical point of Eq.~(\ref{eq:muex.a}) is: this formula is
simple and effective but how it is obtained from the series
Figure~\ref{fig:virial} is not simple.  Part of the complication is
that two additional twists have been interjected, the flat default
model and the continuous approximation.

\subsection{Pressure Denaturation of Proteins}
It is a common view, based upon our current understanding, that
hydrophobic effects provide a nonspecific, cohesive stabilization of
compact protein structures.  However, it has been argued
\cite{Kauzmann:87} that our current understanding of hydrophobic
effects is not consistent with the experimental facts of pressure
denaturation of globular proteins.  The information theory model of
the previous section was applied also to study hydration free energies
and potentials of mean force (pmfs) for two and three hydrophobic
spherical solutes in water as a function of pressure
\cite{Hummer:98}. As is well known, those pmfs exhibit contact and
solvent-separated minima corresponding, respectively, to cases where
the hydrophobic spheres contact each other or where a water molecule
intervenes.  It was found that increasing pressure shifted the free
energy balance of those two cases towards the solvent-separated
circumstance.  This suggested an intercalation mechanism for pressure
denaturation: as the pressure of the liquid is raised, water molecules
are forced into protein structure.  A similar point of view can be
taken of the formation of clathrate hydrates at elevated pressures: at
low pressure hydrophobic effects lead to close contacts and to
clustering of hydrocarbon gases dissolved in water.  Pressure
increases stabilize the crystalline phase that eliminates close
solute contacts.  If attention is focused on the hydrocarbon material
this behavior might seem counter-intuitive because the hydrocarbon
material seems to expand.  But the thermodynamic principle is that
increasing pressure stabilizes the phase of lower volume.  Therefore,
we conclude that the system may be packed more efficiently and have a
lower total volume when water molecules are intercalated into the
hydrocarbon clusters.  These topics will surely be the subject of
further research.

\section{Concluding Comments}
Identification of some generalizations and future directions for these
theories will provide concluding comments.  Firstly, we note that the
generalization of these ideas to treat continuous, rather than only
hard core repulsive, solute-solvent interactions is
known \cite{feature}.

Secondly, we note the importance in the context of the aqueous
solutions of our restriction here to small molecule solutes.  It is
well recognized that treatment of larger solutes requires
consideration of the multiphasic character of these solutions on large
length scales \cite{Stillinger:73,Hummer:96,Hummer:98}. For large
enough hard sphere solutes dissolved in water close to phase
coexistence, the possibility that the solvent will pull away from the
solute surface requires specific attention.  Further subtleties arise
when the solute-solvent interactions are not just repulsive but
include attractive interactions too \cite{Weeks:95,Weeks:97}. These
issues will surely be the subject of further research in the area of
hydrophobic effects.  It seems likely that an appropriately designed
default model should be able to describe such effects in a physical
manner.

Thirdly, we note that these approaches provide some unanticipated
answers \cite{feature} to questions such as ``How is water different
from hydrocarbon liquids as a solvent for nonpolar solutes?'' The
importance of the low and temperature insensitive values of the
isothermal compressibility of liquid water is noteworthy.  However,
such answers are not in the format that is most often intended when
such questions are asked.  Most often such questions solicit
information about particular patterns of solvent structure in the
neighborhood of a hydrophobic solutes.  Some groundwork has been laid
for consideration of those detailed structural issues in a format
consistent with the discussion of this paper \cite{Pratt:98}. Pursuit
of answers about the detailed structural issues and their relevance to
hydrophobic effects will surely be the subject of future research.

Finally, we note again the relevance of these ideas to the classic
problems of packing in liquids.  The importance of these issues is
reflected in the significance of the hard sphere fluid system to our
understanding of liquids.  Hard core model systems may not be directly
realistic.  But it continues to surprise that when attention is
directed to new physical problems, {\em e.g.} the thermodynamics and
structure of glasses or folded proteins, understanding of basic
packing problems is again requested. Such problems surface quite
broadly [66-72]. It would be interesting to see the ordering phase
transitions associated with packing problems, crystallization and
liquid crystallization, analyzed on these bases.  An initial step
along such lines for the hard sphere fluid has been taken
\cite{Crooks:97} but more work is deserved.

We hasten to add that the results so far have not superseded previous
theoretical results.  But this new approach offers the possibility of
better, more physical understandings of packing problems in the
equilibrium statistical mechanics of non-crystalline materials and the
previous theories of them.  This approach {\em has} achieved new
understanding for the problems of primitive hydrophobic effects.  By
exploiting information external to conventional theories, even
simulation data, and by proposing a pattern for utilizing that
information, these approaches begin to respond to Andersen's
\cite{Andersen} request for a `theory of theories.'

\section*{Acknowledgement}  This work was supported by the LDRD
program at Los Alamos.

\end{document}